# Carbon-Based Resistive Memory

Franz Kreupl, Rainer Bruchhaus[+], Petra Majewski, Jan B. Philipp, Ralf Symanczyk, Thomas Happ[*], Christian Arndt[*], Mirko Vogt[*], Roy Zimmermann[*], Axel Buerke[*], Andrew P. Graham[*], Michael Kund

Qimonda AG, [+]Qimonda North America, [*]Qimonda Dresden GmbH & Co. OHG
85579 Neubiberg, Am Campeon 1-12, Germany, franz.kreupl@qimonda.com, phone: +4989600882868, fax: +498960088442868

## Abstract
We propose carbon as new resistive memory material for non-volatile memories and compare three allotropes of carbon, namely carbon nanotubes, graphene-like conductive carbon and insulating carbon for their possible application as resistance-change material in high density non-volatile memories. Repetitive high-speed switching and the potential for multi-level programming have been successfully demonstrated.

## Introduction

Carbon seems to be an ideal material for being electrically manipulated in its conductive state. Carbon exists in many forms, the most prominent ones being the $sp^2$-dominated graphitic form with its high conductivity and the $sp^3$-dominated diamond form which has a low conductivity (1).

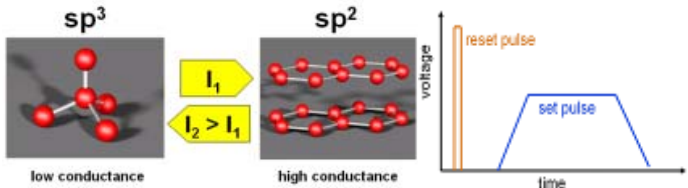

**Figure 1** The principal operation of carbon memory depends on the starting material (insulating carbon or conductive carbon). Insulating carbon can be switched to conductive $sp^2$-rich carbon by inducing electrical breakdown. Conductive carbon can be switched to $sp^3$-rich carbon, by inducing a quenched state with a short high current pulse of sufficient current density, which is in the order of 1GA/cm$^2$.

One main statement of this paper is that one can switch from the low conductive state to the high conductive state and vice versa by applying appropriate current pulses to the carbon element, as indicated in Fig. 1-3 and as is explicitly demonstrated in Fig. 6 and 11. The set pulse which is in the order of tens of nanoseconds forms $sp^2$-rich filaments and the very short reset pulse (~1 ns) results in a disordered, $sp^3$-rich quenched state (1). The two conductance states can be used to represent logical states. In contrast to other switchable memory materials, carbon is of mono-atomic nature and therefore may be scalable to very small feature sizes (even single bonds). One of the key concerns for future high density resistive memories is the current which is needed to switch from one state to the other and which is limited by the current drivability of the select device at reduced dimensions. While carbon will need current densities in the order of GA/cm$^2$, it is scalable below nanometer feature sizes and will limit the overall current to the tens of microampere range – a current value which can be supported by scaled select devices.

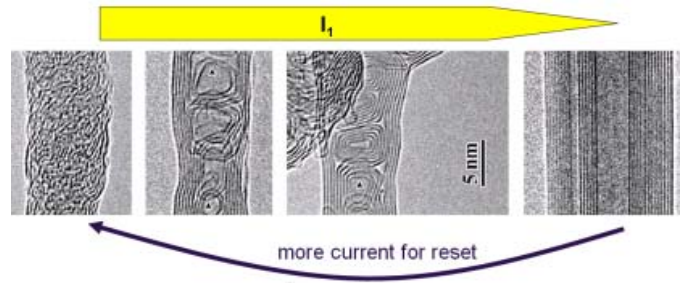

**Figure 2** Visualization of the proposed memory effect in a TEM study (3). A current-source is connected to a disordered carbon fiber in a TEM and the structure change of the fiber is monitored after a current flow. The current switches the disordered fiber (left) to aligned $sp^2$-structures (right image). The resistance of the carbon structure reduces from the left to the right TEM-image by a factor of about 100 (3). TEM images are by courtesy of J.Y. Huang (3).

In addition, the high resilience of carbon will allow for high temperature operations. The energy density $j^2 \cdot \rho \cdot t$ which is applied to a material with resistivity $\rho$, time t of a few ns and current pulse j in the order of GA/cm$^2$ is in the same order of magnitude as it is used in laser ablation techniques (2).

The high temperatures, which will occur, can easily induce structural modifications, as shown in the TEM-image study by Huang et al. (3), shown in Fig. 2, and these structural modification in turn lead to a modified conductance. The time-dependent temperature simulation of a confined carbon filament in Fig. 3 supports this idea even in an integrated structure. The switching of a high resistant state to a low resistant state in carbon is very well known for one-time programmable anti-fuses (4) and the new aspect presented

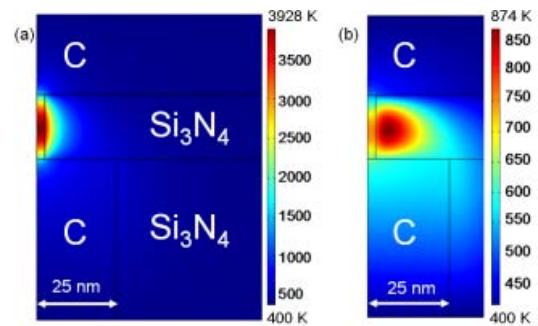

**Figure 3** (a) The temperature distribution in a carbon filament with 5 nm diameter after a 1 ns current pulse with 1.7 GA/cm$^2$ current density has been simulated for a structure similar to Fig. 5. The peak temperature of 3928 K is sufficient to induce modifications in the carbon structure (2). (b) The temperature distribution 0.05 ns after the pulse demonstrates the rapid cooling in the memory structure. Complete thermalization is reached within 0.5 ns.

here is that it is reverse-programmable and that one can switch this low resistant state back to the high resistant state by a short current pulse which induces a quenched, disordered state in the carbon material. There are a lot of options regarding the implementation of a carbon memory cell and the following discussion evaluates three allotropes of carbon regarding their suitability for high density memory applications, which will require operating currents well below 100 μA per cell.

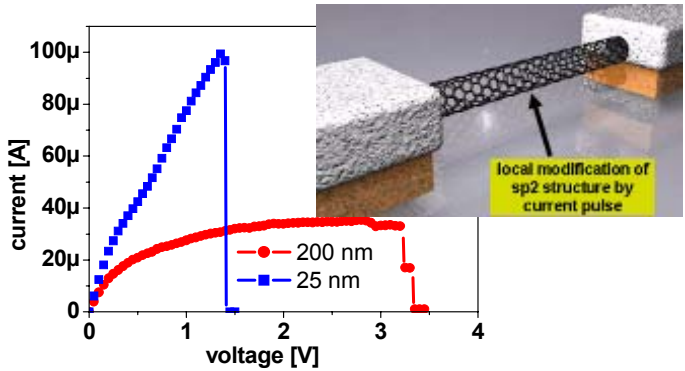

**Figure 4** Experimental determination of the critical (switching) current in single-walled CNTs. For short CNTs (25 nm) there is no current limitation from phonon scattering and the current density exceeds $10^{10}$ A/cm$^2$. Longer (200 nm) CNTs do have current limitation by phonon scattering, but require a higher voltage drop, which needs to be handled by a select device. The required quasi-static power for switching ranges between 100-140 μW.

### Carbon nanotubes

Carbon nanotubes (CNTs) are one sp$^2$-type species of carbon which are known to reduce dramatically the conductance by introducing only a single sp$^3$-bond in their configuration (5). Non-volatile memory effects have also been observed very recently in graphene (6), another sp$^2$-type species of carbon. While CNTs still face the unsolved problem of being placed precisely, they represent one of the smallest available self-assembled carbon structures. The overall current for current-induced switching depends not only on the number of CNT shells but also on the CNT length (7). Due to thermal oscillations of atoms during the current pulse, C–C bonds with high curvature on the CNT wall are susceptible to breaking in the form of non-hexagonal vacancies, which in turn leads to a conductance change. On the other hand, it is very well known that these defective CNTs show self-healing properties if they are subjected to a moderate current flow (3, 8, 9). And very recently, an electron irradiation induced metal to insulator transition in metallic CNTs which is reversible, has been demonstrated (10). The observation can also be attributed to the proposed memory effect although the authors argue that it is induced by trapped charges (10).

The self healing properties require very sharp and short pulses for the preparation of a quenched state, because longer pulses or smooth tails would lead to an annealed low resistant state. Quasi-static I(V) curves, shown in Fig. 4, evaluate the critical current densities in metallic, 25 nm and 200 nm long, single-walled CNTs. Short CNTs require currents in excess of 100μA for switching events and longer CNTs are limited to 20 – 30 μA at the expense of a higher voltage drop, as can be seen in Fig. 4. In summary, the high resilience of CNTs require very high current densities for switching events, but the overall current requirement could be as low as ~ 30 μA for a CNT longer than 200 nm. However, the select device needs to be reliably operated up to 4 Volt in this case.

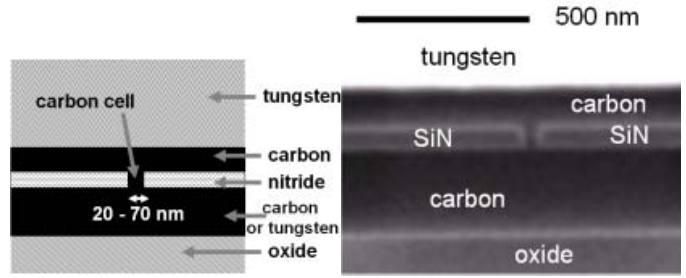

**Figure 5** (a) For the investigation of conductive carbon (CC) samples, the following structure has been fabricated: a 300 nm thick CC or alternatively tungsten bottom contact resides on oxide. Vias with diameters ranging between 20-70 nm have been made in 50 nm thick SiN which has been deposited on the bottom CC and patterned by e-beam lithography and dry etching. The resulting vias have been filled with CC and a CC top contact has been added followed by a tungsten top metallization. (b) FIB-cross-section of a CC memory cell with 30 nm diameter carbon cell and CC bottom and top contact.

### Conductive carbon

We recently have disclosed graphene-like, conductive carbon (CC) with high electrical conductivity as new CMOS-compatible material for interconnect applications (11). This conductive carbon can also be used to realize carbon memory cells. A layout schematic and a cross-section of the fabricated memory cells are shown Fig. 5. The overall structure with carbon top and bottom contact resembles the simulated structure in Fig. 3 except for the smaller carbon cell diameter in the simulated cell. A Shmoo-plot of a 40 nm diameter CC cell is shown in Fig. 6. There is a distinct reaction in the resistance of the cell at 8.5 V pulse. The voltage is quite high, because the cell filament length of 50 nm (thickness of the SiN layer in Fig. 5) requires this voltage to achieve the current density for the preparation of a quenched state.

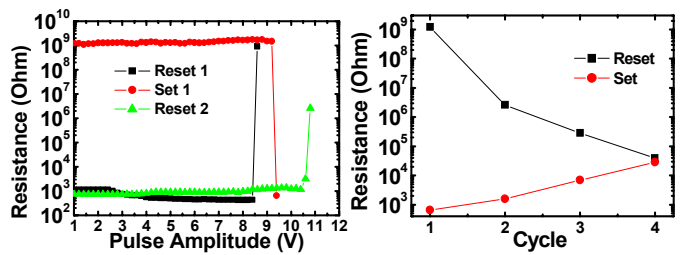

**Figure 6** (a) Shmoo-plot of the resistance of a 40 nm diameter CC memory with tungsten bottom contact for two reset pulses and one set pulse. The pulse width is 5 ns for the reset and 300 ns for the set pulse. No current compliance for the voltage pulse has been used. (b) The resistance values of the memory cell are plotted for the number of switching cycles. The collapse of the $R_{off}/R_{on}$ ratio after 4 cycles is attributed to the missing current compliance, an unadapted pulse length and the formation of a tungsten carbide alloy at the interface of the carbon plug and the tungsten bottom contact. After 4 cycles the cell continues to switch but the $R_{off}/R_{on}$ ratio change is quite small.

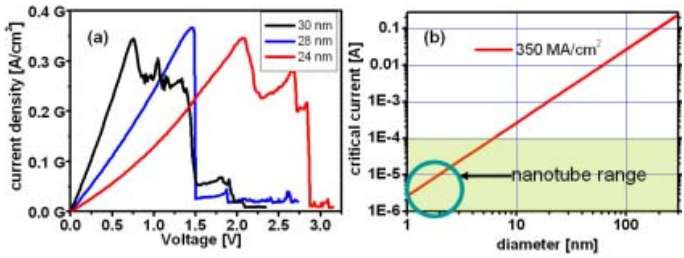

**Figure 7** (a) Critical current densities measured for different conductive carbon cell diameter (24, 28, 30 nm) with CC bottom and top contact. The required current density is in the order of 350 MA/cm$^2$ and the required quasi-static power for switching is 3 mW for the 24 nm wide cell. (b) The determined critical current density can be used to extrapolate feature sizes of the CC memory cell which are compatible with the current drivability of future nano-transistors. Feature sizes well below 10 nm needs to be achieved to bring the required current down to below 100 µA.

The required voltage is expected to scale inversely with carbon cell length. A carbon cell with cell length of 10 nm is therefore expected to switch at voltage of about 1.7 V. In order to determine the critical current density for switching events in CC, cells with varying diameter have been subjected to quasi-static I(V) measurements and the results are depicted in Fig. 7. The cells start to react independently on cell diameter at a current density in the order of 350 MA/cm$^2$. This translates to critical currents of 1.5 mA for a 24 nm diameter cell, which is ways to high for high density applications. The structure sizes which are suitable for low current operations can be estimated in Fig. 7(b) by assuming the measured critical current density as a material constant. From this it can be concluded, that the applicability of conductive carbon is limited to diameter sizes well below 10 nm. However, only the first current pulse should require the high current for the whole carbon diameter for amorphization. Subsequent pulses should utilize only a very thin filament as will be discussed in the next paragraph. Another solution would embrace nano-holes and a few mono-layer thick cladding layer of conductive carbon (11). This would reduce the effect carbon cross-section to that of large diameter single-walled CNTs and reduce the required current accordingly.

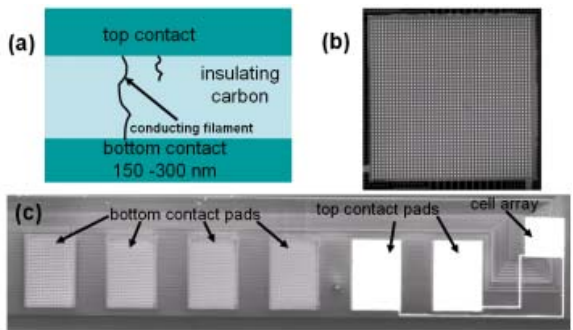

**Figure 8** The experimental setup to study the suitability of insulating carbon (IC). (a) Schematic of the IC memory cell with metallic top and bottom contact and the IC-layer sandwiched in between. A conductive filament will form in the IC layer after electrical breakdown. (b) Top view high voltage SEM revealing continuous top and single bottom plugs of the cell array. (c) SEM image of the cell array and common top and four of the individually addressable bottom contact pads.

## Insulating carbon

Insulating carbon (IC) is available in various forms (12) and has been deposited in this work by plasma-enhanced chemical vapor deposition. Single memory cells have been devised, as shown in Fig. 8. The bottom tungsten plugs which are individually addressable are covered with an 8 to 60 nm thick insulating carbon film. A common top metallization covers the whole cell array. Figure 9 displays quasi-static switching curves with switching currents between 5 and 50µA and switching voltages between 1-3 V depending on the thickness and property of the insulating carbon. Conductive sp$^2$-like filaments are formed in the predominantly sp$^3$-type insulating carbon upon application of voltage pulses (c.f. Fig. 12) (4, 13).

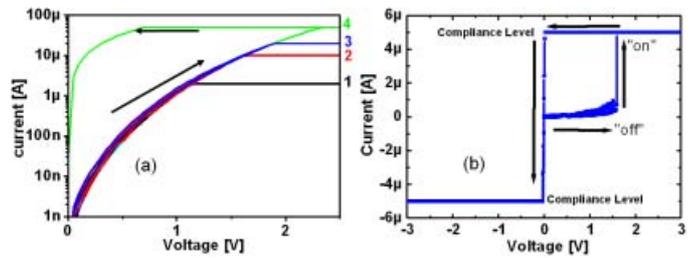

**Figure 9** (a) Quasi-static switching curves for a 350 nm wide contact plug with 8 nm thick insulating carbon. The current compliance was raised in 4 steps from 2µA to 50µA. Burn-in of the conductive filament occurs between 20 and 50 µA. The required switching power is about 50µW. (b) shows switching of an IC cell with 150 nm diameter bottom contact at very low power levels with 1.5 V and 5 µA (P= 7.5 µW).

Fig. 9 (a) shows four voltage sweeps with varying current compliance values ranging from 2, 10 and 20 µA up to 50 µA. The I(V)-curve for the first three compliance settings has the identical trace for up- and down-sweeps, only the fourth sweep leads to a switching effect, which is well known from anti-fuses (4). Reliable switching is demonstrated with only 5 µA current compliance in Fig. 9(b).

Read endurance for the low and high resistance state is shown in Fig. 10 and up to 2.3·10$^{13}$ read cycles have been demonstrated at elevated temperature in the on- and off-state without significant resistance variation. The switching speed depends on the amplitude of the voltage pulse and the measured transition to the low resistance states occurs within

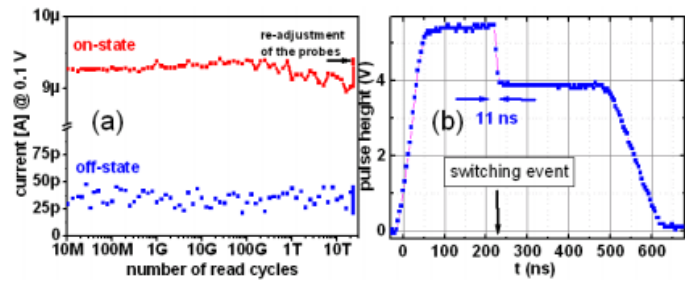

**Figure 10** (a) Read endurance measurement of the on- and off-state of an IC cell at 75°C. The sample has been subjected to 2.3·10$^{13}$ read cycles at 0.1 V. The drift at the end of the measurement in the on-state could be corrected by a re-adjustment of the probe tips. (b) Switching speed measured with a pico-probe upon application of a 5.5 V, 500 ns pulse with 50 ns lead and 100 ns trail. The IC cell reacts after 175 ns and switches to the low resistance state within 11 ns.

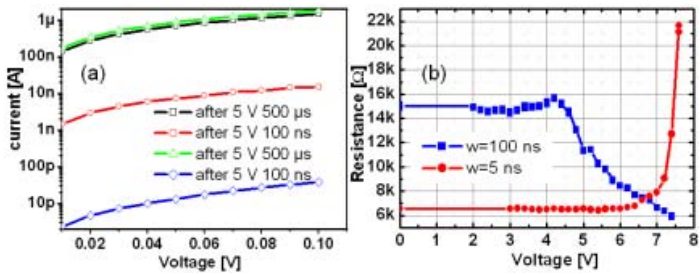

**Figure 11** (a) I(V) curves of an IC memory cell after the application of different voltage pulses. The first pulse switches the cell in a low-resistance state. The second pulse enhances the resistance by a factor of 100. The third pulse recovers the low resistance state and the forth pulse turns the cell off again. (b) Shmoo-plot for the set and reset of the resistance of another IC cell, which is treated with voltage pulses of varying length and height. The resistance can be tuned between 6 kOhm and 22 kOhm by fine-tuning the voltage pulses. Longer pulses lead to an annealing of the $sp^2$-like filaments whereas shorter pulses can induce disorder and lead to a quenched, more resistive state of the carbon filament. This also demonstrates that multi-level programming of a carbon cell is possible by appropriate voltage pulses.

11 ns, as shown in Fig. 10(b). This switching speed is still dominated by the huge capacitance of the contact pads. More importantly, Fig. 11 shows repetitive switching from the high resistance state to the low resistance state upon treatment with pulses and the possibility to program multi-bit states by appropriate pulses. In contrast to charge-induced memory effects in carbon, which require reversed polarity voltage pulses and have limited non-volatility (14, 15), the proposed carbon cell can be operated by unipolar pulses with varying height and length. The current pulse for carbon amorphization needs to be very fast, because the heating of the existing filament can induce further graphitization of the surrounding carbon (3) and can thicken the $sp^2$-filament to very low resistance values which will be difficult to erase. The sustained current densities for metals and carbon differ roughly by a factor of 300, and this can be used to "photograph" the size of the filament, as shown in Fig. 12. The melting of the metal in Fig. 12 pinpoints to a reliability issue at the carbon metal interface and suggests the use of a current spreader made of conductive carbon (like the structure shown in Fig. 5), which can easily withstand the high current densities. In summary, the use of insulating carbon seems to offer the possibility for very low-power operations due to the self-organizing nature of the very thin filament, but the pulse parameters and cell structure still needs to be optimized.

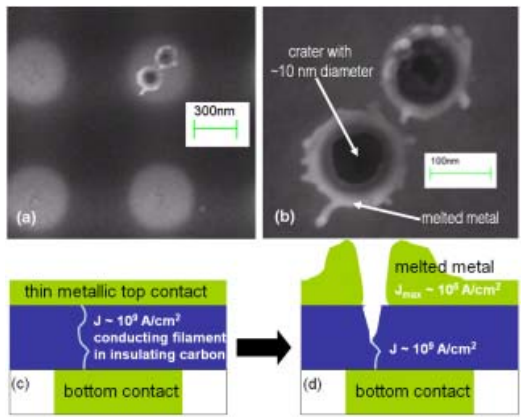

**Figure 12** The approximate size and position of the conductive filament in the IC cell can be visualized by depositing only a very thin (20 nm) top metal layer on the IC film and applying a very high voltage pulse (10 V). Figures (a)–(d) show the 350 nm bottom contact and two spots created by the melted metal at location where the conductive filaments have formed in the IC after the application of two 10 V pulses. The high current density leads to a melting of the very thin metal layer and the hot carbon filament is oxidized and vaporized leaving a crater of the size of the filament, which is about 10 nm in diameter for the high voltage pulse.

## Conclusions

We have proposed carbon as resistive memory material and have evaluated for the first time carbon nanotubes, conductive carbon and insulating carbon for non-volatile memory applications. Other carbon materials like graphene, fullerens and diamond might as well operate on the same principle, but have not been investigated here. Repetitive bistable high-speed switching and the potential for multi-level programming have been successfully demonstrated. It is concluded that the insulating carbon approach seems to be best suited for low power operation as required for future high density applications.

## Acknowledgements
The authors would like to thank R. Thewes, W. Pamler, W.M. Weber, U. Klostermann, K.-D. Ufert, H. Riechert and Stanford University for support.